\documentclass[10pt]{blois07}
\usepackage{graphicx}
\usepackage{epsfig}
\usepackage{cite,./mcite}

\begin{document}
\title{On the behaviour of $R_{pA}$ at high energy\protect\footnote{~~~
    Presented by M.~Kozlov at EDS07.}}
\author{Misha~Kozlov$~^1$,~~Arif~I.~Shoshi$~^1$,~~Bo-Wen Xiao$~^2$}
\institute{$^1$ Fakult{\"a}t f{\"u}r Physik, Universit{\"a}t Bielefeld,
  D-33501 Bielefeld,
  Germany \\
  $^2$ Department of Physics, Columbia University, New York, NY, 10027, USA}
\maketitle

\begin{abstract}
  We discuss the behaviour of $R_{pA}$, the ratio of the unintegrated gluon
  distribution of a nucleus over the unintegrated gluon distribution of a
  proton scaled up by $A^{1/3}$, at high energy and fixed coupling. We show
  that $R_{pA}$ exhibits a rising gluon shadowing with growing rapidity,
  approaching $1/A^{1/3}$ at asymptotic rapidity, which means total gluon
  shadowing due to gluon number fluctuations or Pomeron loops.

\end{abstract}

\section{Introduction}
\label{sec:intro}

We study the ratio of the unintegrated gluon distribution of a nucleus
$\varphi_A(k_{\perp},Y)$ over the unintegrated gluon distribution of a proton
$\varphi_p(k_{\perp},Y)$ scaled up by $A^{1/3}$
\begin{equation}
R_{pA} = \frac{\varphi_{A}\left( k_{\perp },Y\right) }{A^{\frac{1}{3}}\ \varphi_{p}\left(
    k_{\perp },Y\right) } \ .
\label{R_pA}
\end{equation}
This ratio is a measure of the number of particles produced in a
proton-nucleus collision versus the number of particles in proton-proton
collisions times the number of collisions. The transverse
momentum of gluons is denoted by $k_{\perp}$ and the rapidity variable by $Y=\ln(1/x)$.

\begin{figure}[htb]
\setlength{\unitlength}{1.cm}
\par
\begin{center}
\epsfig{file=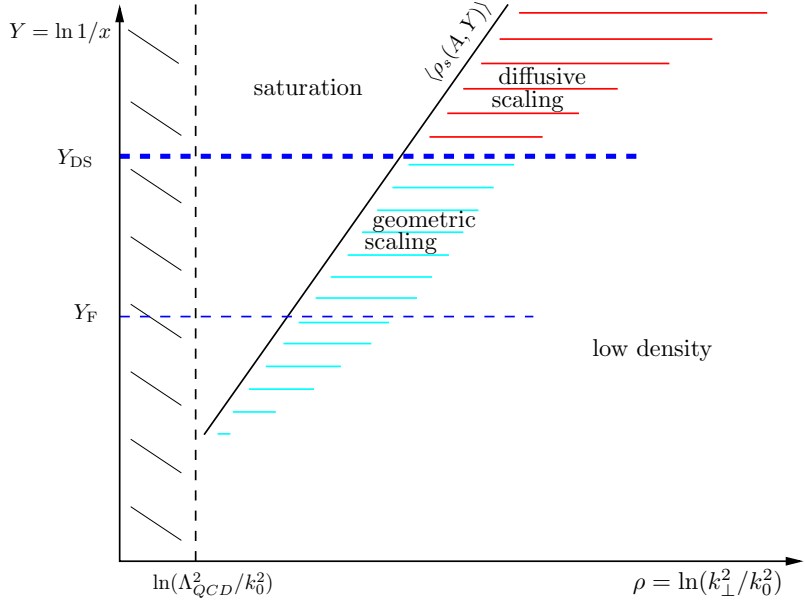, width=11cm} 
\end{center}
\caption{Phase diagram of a highly evolved nucleus/proton.} 
\label{had_wf}
\end{figure}

The ratio $R_{pA}$ has been widely studied~\cite{Iancu:2004bx+X1,
  *Iancu:2004bx+X2, *Iancu:2004bx+X3, *Iancu:2004bx+X4, *Iancu:2004bx+X5,
  *Iancu:2004bx+X6, *Iancu:2004bx+X7,*Iancu:2004bx+X8} in the framework of the
BK-equation~\cite{Balitsky:1995ub+X1, *Balitsky:1995ub+X2,
  *Balitsky:1995ub+X3, *Balitsky:1995ub+X4} which describes the small-$x$
physics in the {\em mean field approximation}. Using the BK-equation one finds
in the geometric scaling regime (transition from high to low gluon density,
see Fig.\ref{had_wf}) in the fixed coupling case that the shape of the
unintegrated gluon distribution of the nucleus and proton as a function of
$k_{\perp}$ is preserved with increasing $Y$, see Fig.\ref{GS}(a), because of
the geometric scaling behaviour
$\varphi_{p,A}(k_{\perp},Y)=\varphi_{p,A}(k_{\perp}^2/Q^2_s(Y))$, and
therefore the leading contribution to the ratio $R_{pA}$ is basically
$k_{\perp}$ and $Y$ independent, scaling with the atomic number $A$
as~\cite{Mueller:2003bz,Iancu:2004bx}
\begin{equation}
R_{pA} \simeq \frac{1}{A^{\frac{1}{3}(1-\gamma_{_0})}} \ ,
\label{eq:RGS}
\end{equation}
where $\gamma_{_0}=0.6275$. This means that gluons inside the nucleus and
proton are somewhat shadowed since $\varphi_A/\varphi_p = A^{\gamma_{_0}/3}$
lies between total ($\varphi_A/\varphi_p=1$) and zero
($\varphi_A/\varphi_p=A^{1/3}$) gluon shadowing. The {\em partial gluon
  shadowing} comes from the anomalous behaviour of the unintegrated gluon
distributions which stems from the BFKL evolution. The partial gluon shadowing 
may explain why particle production in heavy ion collisions scales, roughly, 
like $N_{part}$~\cite{Kharzeev:2002pc}.

Over the last few years, it has been understood how to deal with small-$x$
physics at high energy {\em beyond the mean field approximation}, i.e., beyond
the BK~\cite{Balitsky:1995ub+X1, *Balitsky:1995ub+X2, *Balitsky:1995ub+X3,
  *Balitsky:1995ub+X4} and JIMWLK~\cite{Jalilian-Marian:1997jx+X1,
  *Jalilian-Marian:1997jx+X2, *Jalilian-Marian:1997jx+X3,
  *Jalilian-Marian:1997jx+X4, *Jalilian-Marian:1997jx+X5} equations. We have
learned how to account for the elements missed in the mean field evolution,
such as the descreteness and fluctuations of gluon
numbers~\cite{Mueller:2004sea,Iancu:2004es} or the Pomeron
loops~~\cite{Mueller:2005ut+X1, *Mueller:2005ut+X2, *Mueller:2005ut+X3,
  *Mueller:2005ut+X4, *Mueller:2005ut+X5}.  The main result as a consequence
of the above is the emerging of a new scaling behaviour for the
dipole-hadron/nucleus scattering amplitude at high
rapidities~\cite{Mueller:2004sea,Iancu:2004es}, the so-called diffusive
scaling.  This is different from the geometric scaling behaviour which is the
hallmark of the "mean-field" evolution equations (JIMWLK and BK equations).
The effects of fluctuations on the scattering
amplitude~\cite{Brunet:2005bz+X1, *Brunet:2005bz+X2, *Brunet:2005bz+X3,
  *Brunet:2005bz+X4, *Brunet:2005bz+X5, *Brunet:2005bz+X6, *Brunet:2005bz+X7,
  *Brunet:2005bz+X8, *Brunet:2005bz+X9, *Brunet:2005bz+X10,
  *Brunet:2005bz+X11, *Brunet:2005bz+X12, *Brunet:2005bz+X13,
  *Brunet:2005bz+X14}, the diffractive scattering processes\cite{Hatta:2006hs,
  Kovner:2006ge+X1, *Kovner:2006ge+X2, *Kovner:2006ge+X3} and forward gluon
production in hadronic scattering processes~\cite{Iancu:2006uc,Kovner:2006wr}
has been studied so far. In this work we show how the behaviour of $R_{pA}$ as
a function of $k_{\perp}$ and $Y$ in the fixed coupling case is completely
changed due the effects of gluon number fluctuations or Pomeron loops at high
rapidity~\cite{KSX_1,*KSX_2}.

%
\section{$R_{pA}$ ratio in the diffusive scaling regime}
\label{sec:Rpa}
According to the statistical physics/high energy QCD
correspondence~\cite{Iancu:2004es} the influence of fluctuations on the
unintegrated gluon distribution of a nucleus/proton is as follows: Starting
with an intial gluon distribution of the nucleus/proton at zero rapidity, the
stochastic evolution generates an ensamble of distributions at rapidity $Y$,
where the individual distributions seen by a probe typically have different
saturation momenta and correspond to different events in an experiment. To
include gluon number fluctuations one has to average over all individual
events, 
\begin{equation}
\langle \varphi_{p,A}(\rho-\rho_s)\rangle = \int d\rho_s\ \varphi_{p,A}(\rho-\rho_s) \
P(\rho_s-\langle\rho_s\rangle) \ ,
\label{av_gd}
\end{equation}
where $\varphi_{p,A}(\rho-\rho_s)$ is the distribution for a single event with
$\rho = \ln(k_{\perp}^2/k_0^2)$ and
$ P(\rho_s-\langle\rho_s\rangle)$ the probability distribution of the
logarithm of the saturation momentum, $\rho_s(Y)=\ln(Q^2_s(Y)/k_0^2)$, which
is argued to have a Gaussian form~\cite{Marquet:2006xm},
\begin{equation}
  P(\rho _{s})\simeq \frac{1}{\sqrt{2\pi \sigma ^{2}}}\exp \left[ -\frac{%
      \left( \rho _{s}-\langle \rho _{s}\rangle \right) ^{2}}{2\sigma ^{2}}\right]
  \quad \mbox{for} \quad \rho-\rho_s \ll \gamma_c^2 \sigma^2 \ ,
\label{proba_gauss}
\end{equation}
with the dispersion
\begin{equation}
\sigma^2 = \langle \rho_s^2 \rangle- \langle \rho_s \rangle^2 = D Y.  
\label{sigma_v}
\end{equation}
The main consequence of fluctuations is the replacement of the geometric
scaling, $\varphi_{p,A}(k_{\perp},Y)=\varphi_{p,A}(k^2_{\perp}/Q_s^2(Y))$, by
a new scaling, the diffusive scaling~\cite{Mueller:2004sea,Iancu:2004es},
namely, $\langle \varphi_{p,A}(k_{\perp},Y) \rangle$ is a function of another
scaling variable ($\langle Q_s \rangle$ is the average saturatin momentum), 
\begin{equation}
\langle \varphi_{p,A}(k_{\perp},Y)\rangle =
F_{p,A}\left(\frac{\ln(k^2_{\perp}/\langle Q_s(Y)\rangle^2)}{DY}\right) \ .
\end{equation}
The diffusive scaling, see Fig.~\ref{had_wf}, sets in when the dispersion of
the different events is large, $\sigma^2 = \langle \rho_s^2 \rangle- \langle
\rho_s \rangle^2 = D Y \gg 1$, i.e., $Y \gg Y_{DS} =1/D$ where $D$ is the
diffusion coefficient, and is valid in the region $\sigma \ll
\ln(k^2_{\perp}/\langle Q_s(Y)\rangle^2) \ll \gamma_{_0}\, \sigma^2$. 

\begin{figure}[htb]
\setlength{\unitlength}{1.cm}
\par
\begin{center}
\epsfig{file=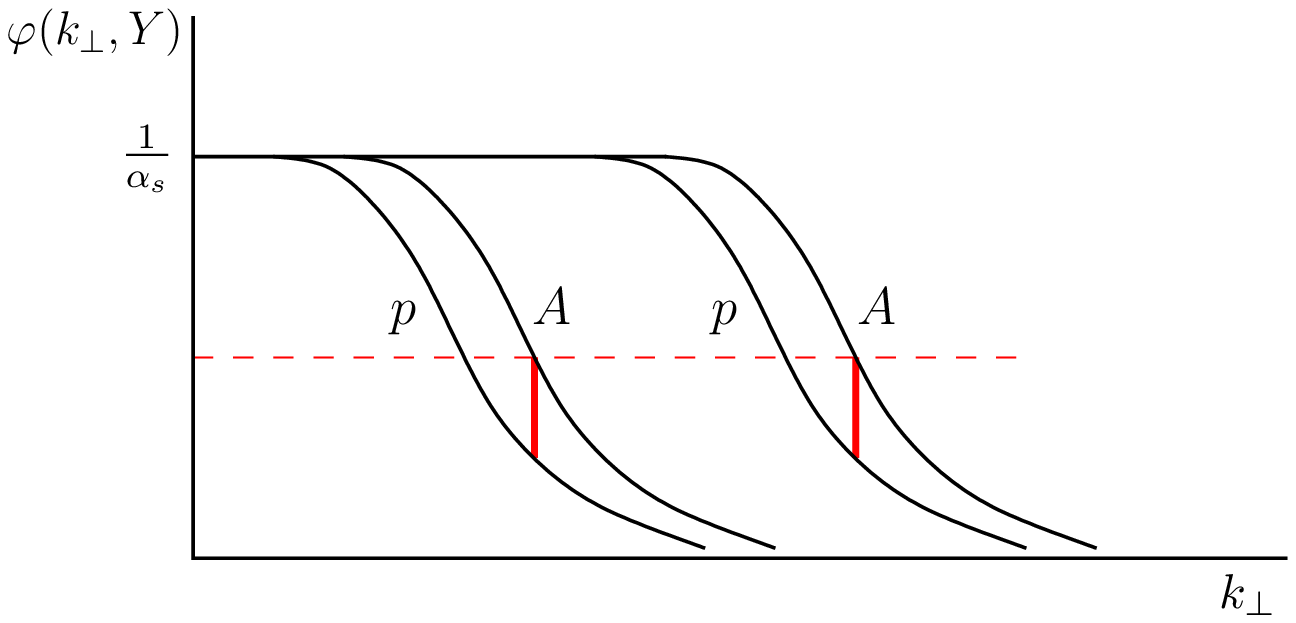, width=7.0cm} ~~
\epsfig{file=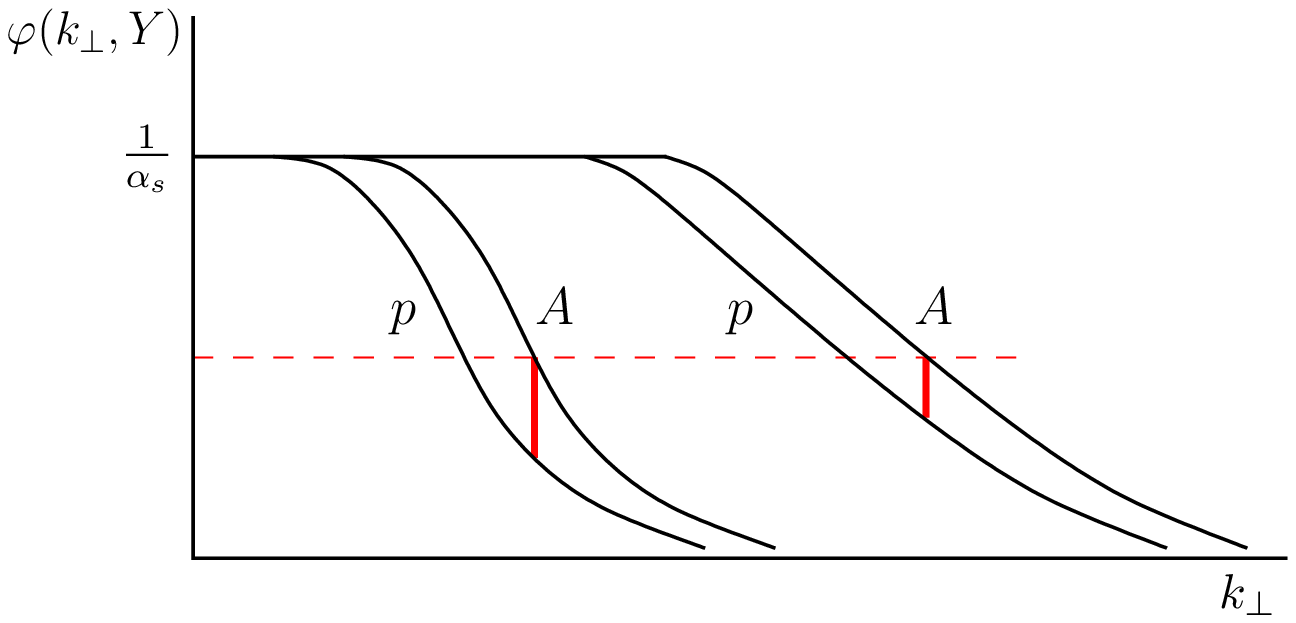, width=7.0cm} \\
\hspace*{0.3cm} (a) \hspace*{9cm} (b)
\end{center}
\caption{The qualitative behaviour of the unintegrated gluon distribution of a
  nucleus (A) and a proton (p) at two different rapidities in the geometric
  scaling regime (a) and diffusive scaling regime (b). }
\label{GS}
\end{figure}

The diffusive scaling means that the shape of the unintegrated gluon
distribution of the nucleus/proton changes with increasing $Y$ because of the
additional $DY$ dependence as compared with the geometric scaling. The shape
becomes flatter and flatter with increasing rapidity $Y$, as shown in
Fig.\ref{GS}(b), in contrast to the preserved shape in the geometric scaling
regime shown in Fig.\ref{GS}(a). This flattening will lead to a new phenomenon
for $R_{pA}$ as discussed below.

Using Eq.(\ref{av_gd}) for the averaging over all events and the result from
the BK-equation for the the single event distribution one obtains~\cite{KSX_1}
for the ratio
\begin{equation}
R_{pA} \simeq \frac{1}{A^{\frac{1}{3}(1 - \frac{\Delta\rho_s}{2\sigma^2})}}\ 
\left[\frac{k_{\perp}^2}{\langle
  Q^2_s(A,y)\rangle}\right]^{\frac{\Delta\rho_s}{\sigma^2}}
\label{eq:RPA1}
\end{equation}
with the difference between the average saturation
lines of the nucleus and the proton  
\begin{equation}
\Delta \rho_s \; \equiv \;  \langle \rho_s(A,Y)\rangle -  \langle \rho_s(p,Y)\rangle
\; = \; \ln \frac{\langle Q_s(A,Y)\rangle^2}{\langle Q_s(p,Y)\rangle^2} \ 
\label{del_rho}
\end{equation}
where $\langle Q_s(A,Y)\rangle$ ($\langle Q_s(p,Y)\rangle$) is the average
saturation momentum of the nucleus (proton).  The difference $\Delta \rho_s$
is fixed by the inital conditions for the average saturation momenta of the
nucleus and proton and is $Y$-indipendent in the fixed coupling case. For
example, using the known assumption $\langle Q_s(A)\rangle^2 = A^{1/3}\,\langle
Q_s(p)\rangle^2$ one obtains $\Delta \rho_s = \ln A^{1/3}$.

\begin{figure}[htb]
\setlength{\unitlength}{1.cm}
\par
\begin{center}
\epsfig{file=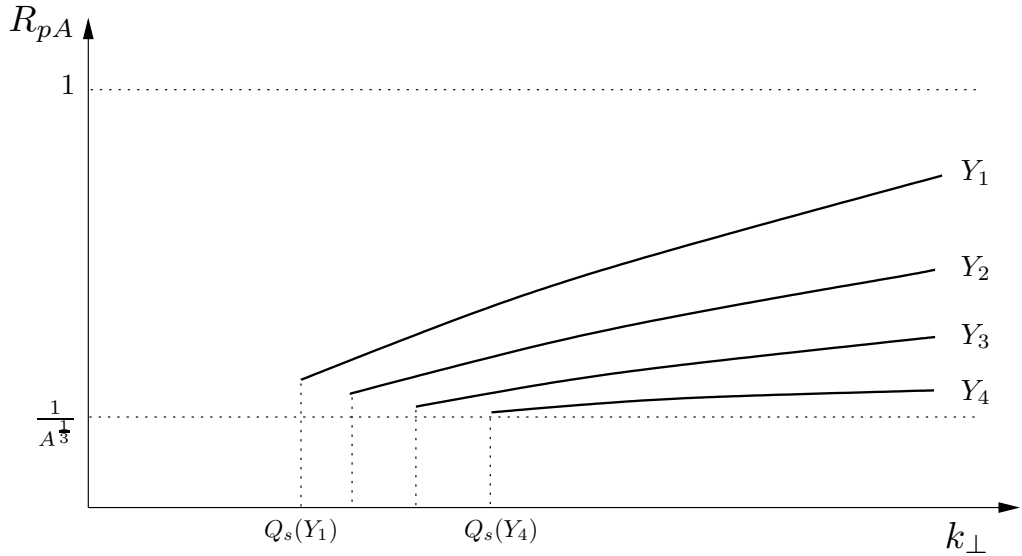, width=14cm}
\end{center}
\caption{The qualitative behaviour of the ratio $R_{pA}$ as a function of $k_{\perp}$ at four different
  rapidities, $Y_1 \leq Y_2 \leq Y_3 \leq Y_4$, in the diffusive scaling regime.
  $R_{pA}$ is always smaller than one for values of $k_{\perp}$ in the diffusive
  scaling regime.}
\label{R_DS}
\end{figure}

The ratio $R_{pA}$ in Eq.~(\ref{eq:RPA1}) shows the following very different
features as compared with the ratio in the geometric scaling regime given in 
Eq.~(\ref{eq:RGS}):
\begin{itemize}
\item In the diffusive scaling regime where $k^2_{\perp}$ is close to $\langle
  Q_s^2(A,Y)\rangle$, the gluon shadowing characterized by
  $A^{\frac{1}{3}(\frac{\Delta\rho_s}{2\sigma^2}-1)}$ increases as the
  rapidity grows (at fixed $A$ or $\Delta\rho_s$) because of $\sigma^2 = DY$.
  At asymptotic rapidity one obtains {\em total gluon shadowing},
  $R_{pA}=A^{1/3}$, which means that the unintegrated gluon distribution of
  the nucleus and that of the proton become the same in the diffusive scaling
  regime at $Y \to \infty$. The phenomenon of total gluon shadowing is
  universal since it does not depend on the initial conditions ($\Delta\rho_s$).

\vskip 3mm

Total gluon shadowing is an effect of gluon number fluctuations (or Pomeron
loops) since fluctuations make the unintegrated gluon distributions of the
nucleus and of the proton flatter and flatter~\cite{Iancu:2004es} and their
ratio closer and closer to $1$ (at fixed $\Delta\rho_s$) with rising rapidity,
as shown in Fig.\ref{GS}(b). Total gluon shadowing is not possible in the
geometric scaling regime in the fixed coupling case since the shapes of the
gluon distributions of the nucleus and of the proton remain the same with
increasing $Y$ giving a constant ratio unequal one, as shown in
Fig.\ref{GS}(a). In the absence of fluctuations one can expect only partial
gluon shadowing, see Eq.~(\ref{eq:RGS}), in the fixed coupling case.

\item The ratio $R_{pA}$ increases with rising $k^2_{\perp}$ within the
  diffusive scaling region. Since the exponent $\Delta\rho_s/\sigma^2$
  decreases with rapidity, the slope of $R_{pA}$ as a function of
  $k_{\perp}^2$ becomes smaller with growing $Y$. The result for $R_{pA}$ in
  the diffusive scaling regime in Eq.(\ref{R_DS}) is very different from the
  result obtained in the mean field approximation given in Eq.~(\ref{eq:RGS}),
  where gluon number fluctuations are not included, which is basically
  $k_{\perp}$ and $Y$-independent.
\end{itemize}
\vskip 3mm

The qualitative behaviour of $R_{pA}$ as a function of $k_{\perp}$ at four
different rapidities, $Y_1 \leq Y_2 \leq Y_3 \leq Y_4$, in the diffusive
scaling regime and for a fixed coupling is shown in Fig.~\ref{R_DS}. Note that
$R_{pA}$ is always smaller than one for values of $k_{\perp}$ in the diffusive
scaling regime.


The above effects of fluctuations on $R_{pA}$ are valid in the fixed coupling
case and at very large energy. It isn't clear yet whether the energy at LHC is
high enough for them to become important. Recently, while in
Ref.~\cite{Kozlov:2007wm} a possible evidence of gluon number fluctuations in
the HERA data has been found, in Ref.~\cite{Dumitru:2007ew}, using a toy
model, it has been argued that in case of a running coupling fluctuations can
be neglected in the range of HERA and LHC energies. See also Refs.~\cite{Mueller:2004sea,Beuf:2007qa} for
more studies on running coupling plus fluctuation effects.

Moreover, the running of the coupling~\cite{Albacete:2007yr+X1,
  *Albacete:2007yr+X2, *Albacete:2007yr+X3, *Albacete:2007yr+X4,
  *Albacete:2007yr+X5,*Albacete:2007yr+X6} may become more important than the effect of gluon
number fluctuations~\cite{Dumitru:2007ew}. In case of a running coupling, the
gluon shadowing increases with rising rapidity in the geometric scaling
regime~\cite{Iancu:2004bx+X1, *Iancu:2004bx+X2, *Iancu:2004bx+X3,
  *Iancu:2004bx+X4, *Iancu:2004bx+X5, *Iancu:2004bx+X6, *Iancu:2004bx+X7,*Iancu:2004bx+X7}, as
opposed to the (roughly) fixed value (partial shadowing) in the fixed-coupling
case, and would lead to total gluon shadowing~\cite{Iancu:2004bx} at very high
rapidities even if fluctuations were absent. In case fluctuations are
important at LHC energy, in addition to the theoretically interesting
consequences of fluctuations on $R_{pA}$, the features of $R_{pA}$ worked out
here, as the increase of the gluon shadowing and the decrease as a function of
the gluon momentum with rising rapidity, may be viewed as signatures for
fluctuation effects in the LHC data. More work remains to be done in
order to clarify how important fluctuation or running coupling effects are at
given energy, e.g., at LHC energy. An extension of this work by the
running coupling may help to clarify some of the open questions. 

\vskip 3mm

\leftline{\bf Acknowledgments}
\vskip 2mm
\noindent A. Sh. and M. K.  acknowledge financial support by the Deutsche
Forschungsgemeinschaft under contract Sh 92/2-1.

\begin{footnotesize}
\bibliographystyle{blois07} 
{\raggedright
\bibliography{blois07}
}
\end{footnotesize}

\end{document}